\documentclass{article}
\usepackage{graphicx} 
\usepackage{tikz}
\usepackage{hyperref}
\usepackage{wrapfig}
\usepackage{graphicx,amsmath,amsfonts,amssymb,amsthm,bbold,dsfont,authblk,cite,xcolor}
\usepackage{fullpage}
\usepackage[mathscr]{euscript}
\usepackage[ruled,vlined]{algorithm2e}
\usepackage{subcaption}

\definecolor{DarkGreen}{rgb}{0.1,0.5,0.1}
\definecolor{DarkBlue}{rgb}{0.1,0.1,0.5}
\hypersetup{
    unicode=false,          
    pdftoolbar=true,        
    pdfmenubar=true,        
    pdffitwindow=false,      
    pdfnewwindow=true,      
    colorlinks=true,       
    linkcolor=DarkBlue,          
    citecolor=DarkGreen,        
    filecolor=DarkGreen,      
    urlcolor=DarkBlue,          
    pdftitle={},
    pdfauthor={},
}

\newcommand{\sect}[1]{\S\ref{#1}}                 

\newtheorem{thm}{Theorem}[section]

\title{Tesseract: A Search-Based Decoder for Quantum Error Correction}
\author{Laleh Aghababaie Beni, Oscar Higgott, Noah Shutty}
\affil{Google Quantum AI, Venice, CA, 90291}
\date{March 1, 2025}

\begin{document}

\maketitle

\begin{abstract}
Tesseract is a Most-Likely Error decoder designed for low-density-parity-check quantum error-correcting codes. Tesseract conducts a search through a graph on the set of all subsets of errors to find the lowest cost subset of errors consistent with the input syndrome.
Although this graph is exponentially large, the search can be made efficient in practice for random errors using $A^*$ search technique along with a few pruning heuristics.
We show through benchmark circuits for surface, color, and bivariate-bicycle codes that Tesseract is significantly faster than integer programming-based decoders while retaining comparable accuracy at moderate physical error rates.
We also find that Tesseract can decode transversal CNOT protocols for surface codes on neutral atom quantum computers.
Finally, we compare surface code and bivariate bicycle code circuits, finding that the [[144,12,12]] bivariate bicycle code is $14\times$ to $19
\times$ more efficient than surface codes using our most-likely error decoding, whereas using correlated matching and BP+OSD decoders would have implied only a $10\times$ improvement.
Assuming instead that long-range couplers are $10\times$ noisier, the improvement drops to around $4\times$ using Tesseract or $2\times$ using correlated matching and BP+OSD.
\end{abstract}
\section{Introduction}
The implementation of quantum error correction (QEC) requires fast and accurate decoders to achieve low logical error rates.
Decoding is an NP-hard optimization problem in the worst case but a long and beautiful line of work has provided a variety of partial solutions that apply to specific codes.
An important class of QEC codes are the low-density parity check (LDPC) codes.
A full review of decoding algorithms for quantum LDPC codes is out of scope but we refer e.g. to the survey \cite{demarti2024decoding}.
Many approaches start with an algorithm that has a polynomial runtime and use heuristics to improve the accuracy.
The Tesseract decoder takes a different approach.
We begin with an exponential-time algorithm that always identifies the most-likely error and use heuristics to make it faster.

\noindent {\bf Organization:} In \sect{sec:notation} we define the framework in which the Tesseract decoding algorithm operates. In \sect{sec:algorithmAndOptimizations} we introduce the algorithm and explain the most important optimizations that make it fast in practice.
In \sect{sec:results} we apply Tesseract and the integer program decoder to several circuits under SI1000 noise~\cite{Gidney2022benchmarkingplanar}: rotated surface code memories \cite{dennis2002topological}, color code memories \cite{bombin2006topological} with the superdense circuit schedule \cite{baireuther2019neural,gidney2023new}, transversal CNOT operations between rotated surface codes \cite{cain2024correlated}, and bivariate bicycle code memories \cite{bravyi2024high}. For the bicycle codes, we also compare with BP+OSD \cite{panteleev2021degenerate}. Finally, we compare the surface and bivariate bicycle codes to each other using these near-optimal decoders.

\section{Notation}\label{sec:notation}
There are various equivalent sets of terminology for discussing binary linear codes.
For efficiency of exposition and to best mirror the source code\footnote{Open-source code for the Tesseract decoder and the Integer Program based decoder available  at \href{https://github.com/quantumlib/tesseract-decoder}{github.com/quantumlib/tesseract-decoder}.}, we will use the terminology associated with the \texttt{DetectorErrorModel} in Stim \cite{stim}.
Throughout, we will let $\mathscr{E} = \{e_1, e_2, \ldots, e_N\}$ denote the set of errors and $\mathscr{D} = \{d_1, d_2, \ldots, d_K\}$
denote the set of detectors of an error model.
\footnote{\
Note that ``error" and ``detector" correspond to existing concepts in classical binary linear codes.
In a classical binary linear code with parity check matrix $H\in \{0,1\}^{K\times N}$, the errors would correspond to `codeword bits' i.e. columns of $H$ and the detectors would correspond to `parity checks' i.e. rows of $H$.
Although it is not our primary motivation here, we point out the Tesseract decoder can also be applied as a maximum-likelihood decoder for classical binary linear codes.
}
We let $p: \mathscr{E}\rightarrow (0, 1/2]$ be an assignment of the probability of each error.\footnote{WLOG we may assume that all errors occur with a nonzero probability that is at most $1/2$ \cite{higgott2025sparse}.}
We will let $w: \mathscr{E}\rightarrow \mathbb{R}^+$ be the function $w(e) = -\log {(\frac{p}{1-p})}$.
We assume that the error $e_i$ is activated independently with probability $c(e_i)$.
For a finite set $S$, we let $2^S$ denote the power set of $S$.
For finite sets $S, T$, we let $S\oplus T = (S\cup T) \setminus (S\cap T)$.
We let $E: \mathscr{D}\rightarrow 2^\mathscr{E}$ be the function such that $E(d)$ is the set of errors that activate detector $d$.
We let $D: \mathscr{E}\rightarrow 2^\mathscr{D}$ be the function such that $D(e)$ is the set of detectors activated by error $e$.
We will abuse notation: if $F\subset \mathscr{E}$ is a set of errors, we will let $D(F) := \bigoplus_{e\in F}D(e)$.
We will also fix a set $\mathscr{S}\subset \mathscr{D}$ which is the set of ``activated detectors" received as input to the decoder.
This is often called the set of ``detection events".
The Most-Likely Error problem is to find
\begin{equation}
    \mathrm{argmin}_{\substack{F\subset \mathscr{E} \\ D(F) = \mathscr{S}}} w(F) \label{eq:MOLE}
\end{equation}
The problem~\eqref{eq:MOLE} may be formulated as an integer program \cite{landahl2011fault,cain2024correlated,lacroix2024scaling} and solved exactly. We use this method as a baseline to compare the Tesseract decoder against. Specifically, we use the High Performance Optimization
Software (HiGHS) package to solve the integer program \cite{huangfu2018parallelizing}.

Now we define some objects required to explain the Tesseract decoding algorithm.
First, we let $G = (2^{\mathscr{E}}, T, w)$ denote a weighted directed graph on the powerset of errors with directed edge set $T$.
We define $T$ in a rather obtuse way to allow us to parametrize $G$ by a choice of predicate $P: 2^\mathscr{E}\times 2^\mathscr{E}\rightarrow \{0, 1\}$:
\begin{equation}
    T = \left\{ (F, F') : (|F'| = |F| + 1) \land (F\subset F') \land P(F, F') \right\}
\end{equation}

In words, the edges out from a vertex $F\in 2^{\mathscr{E}}$ simply correspond to adding one new error to the set $F$.
But not just any error can be added -- only errors that satisfy the predicate $P(F, F')$.

\begin{wrapfigure}{r}{0.5\linewidth}
\centering
\begin{tikzpicture}[scale=1.5, every node/.style={rectangle, draw, fill=white, inner sep=4pt}]
\node (000) at (0, 0) {$\emptyset$};
\node (001) at (-1, 1) {$\{3\}$};
\node (010) at (0, 1) {$\{2\}$};
\node (100) at (1, 1) {$\{1\}$};
\node (011) at (-1, 2) {$\{2, 3\}$};
\node (101) at (0, 2) {$\{1, 3\}$};
\node (110) at (1, 2) {$\{1, 2\}$};
\node (111) at (0, 3) {$\{1, 2, 3\}$};

\draw[->] (000) -- (001);
\draw[->] (000) -- (010);
\draw[->] (000) -- (100);
\draw[->] (001) -- (011);
\draw[->] (001) -- (101);
\draw[->] (010) -- (011);
\draw[->] (010) -- (110);
\draw[->] (100) -- (101);
\draw[->] (100) -- (110);
\draw[->] (011) -- (111);
\draw[->] (101) -- (111);
\draw[->] (110) -- (111);

\end{tikzpicture}
    \caption{The graph $G$ on all error sets, if $|\mathscr{E}| = 3$ and the predicate $P(F, F')$ is always $1$. Weights of the edges are not shown.}
    \label{fig:G4}
\end{wrapfigure}
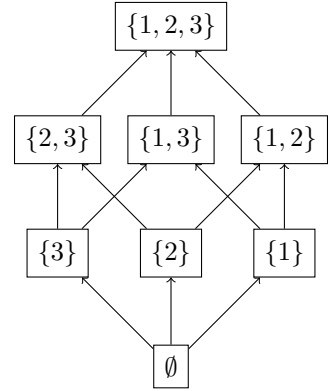

For now, the reader may imagine that the predicate always evaluates to $1$, so that the edges of $G$ just correspond to adding any single error.
An illustration of the resulting $G$ for this case when $N = 3$ is shown in Figure~\ref{fig:G4}.
But later on in \sect{sec:algorithmAndOptimizations} we will describe how the predicate can be made more restrictive so that $G$ has a lower degree, which can be used to improve the efficiency of the implementation.

Recall that the weight $w(e)$ was defined before as the cost of a single error $e$. But we abuse notation now to let the weight of an edge $(F, F')\in T$ be denoted by
$w((F, F')) = w(e)$ where $F'\setminus F = \{e\}$.
We apologize to the reader for these abuses of notation.

We refer to the empty set as the START node: $\mathrm{START} := \emptyset$.
We define the set of EXIT nodes as the set of sets of errors consistent with the syndrome:
\begin{equation}
    \mathrm{EXIT} = \{F\in 2^{\mathscr{E}} : D(F) = \mathscr{S}\}
\end{equation}

It is immediate from the definition of $G$ that it is an acyclic graph and so the set of paths $P_G$ through $G$ is finite:
\begin{equation}
    P_G := \{(F_1, \ldots, F_k) : k\in \mathbb{N} \text{ and } (F_i, F_{i+1})\in T \, \forall i\in \{1, \ldots, k-1\}\}.
\end{equation}
We observe that a path exists from $F$ to $F'$ in $P_G$ if and only if $F'\supset F$.
In that case, all paths from $F$ to $F'$ have the same total edge cost, which we denote $d_G(F, F')$:
\begin{equation}
    d_G(F, F') := \begin{cases}
    \sum_{e\in F'\setminus F}w(e) & F'\supset F\\
    \infty & \text{otherwise}
\end{cases}.
\end{equation}
We will abuse notation by allowing, when $S\subset 2^{\mathscr{E}}$,
\begin{equation}
    d_G(F, S) = \min_{F'\in S}d_G(F, F').
\end{equation}
\section{Decoding as Optimized Path-Finding in $G$}\label{sec:algorithmAndOptimizations}
We have defined the graph $G$ because the Most-Likely Error problem is equivalent to a pathfinding problem on $G$, as formalized in Theorem~\ref{thm:MoleIsPathfinding}.
\begin{thm}\label{thm:MoleIsPathfinding}
The Most-Likely Error problem of eq~\eqref{eq:MOLE} is the Shortest Path problem in $G$, i.e.:
\begin{equation}\label{eq:MoleIsPathfinding}
    \mathrm{argmin}_{\substack{F\subset \mathscr{E} \\ D(F) = \mathscr{S}}} w(F) = \mathrm{argmin}_{\mathrm{END}\in \mathrm{EXIT}} d_G(\mathrm{START}, \mathrm{END})
\end{equation}
\begin{proof}
Observe that
\begin{equation*}
d_G(\mathrm{START}, \mathrm{END}) = \sum_{e\in \mathrm{END}}w(e).
\end{equation*}
Recall that by definition,
\begin{equation*}
\mathrm{EXIT} = \{F\subset \mathscr{E} : D(F) = \mathscr{S} \}.
\end{equation*}
Substituting these definitions, we are left with eq~\eqref{eq:MoleIsPathfinding}.
\end{proof}
\end{thm}
This means that we can apply pathfinding algorithms. Dijkstra's algorithm \cite{dijkstra2022note} when applied to this graph, amounts to a form of brute-force search where we iterate over the sets of errors in order of increasing cost until we reach a valid decoding $\mathrm{END}\in \mathrm{EXIT}$ which is an early stop to the traditional Dijkstra's algorithm.  
Tesseract is able to go much faster than this by exploiting the A* pathfinding algorithm~\cite{hart1968formal}, as explained below.
We will now explain the most important optimizations used to make Tesseract decode quickly in practice.

\noindent{\bf Pruning the graph:} As explained in \sect{sec:notation}, we prune the graph $G$ by choosing a predicate $P: 2^\mathscr{E}\times 2^\mathscr{E}\rightarrow \{0, 1\}$.
If this predicate is a constant function that always evaluates to 1, there are $|F|!$ different paths from $\emptyset$ to $F\in 2^{\mathscr{E}}$. This redundancy is expensive because the search algorithm has to explore more nodes of the graph. We can eliminate the redundancy by canonicalizing the order in which errors are added. We do so using the predicate $P_{\mathrm{T}}$ defined in Algorithm~\ref{alg:prune}.
However, it may be easier to simply explain in words how this predicate works. We start with the residual syndrome at node $F$, which we denote $x$.
The main idea is that we limit the errors $e\in \mathscr{E}$ that can be added to $F$ to be those incident to the lowest index activated detector in $x$. To further limit the search space, we don't allow to add any ``forbidden" errors. We forbid errors in one of two ways:
\begin{enumerate}
    \item \texttt{GetForbiddenErrorsByPrecedence}($F$)  returns the set of errors which we could have added at a previous state transition on the path to $F$, if we chose to add a higher-index error instead.
    \item \texttt{GetForbiddenErrorsAtMostTwo}($F$)  returns the set of errors such that adding the error $e$ would result in more than 2 errors incident to a single detector.
\end{enumerate}
It is easy to see that both routines \texttt{GetForbiddenErrorsPrecedence} and  \texttt{GetForbiddenErrorsAtMostTwo} make the graph $G$ into a tree, which simplifies the traversal as we no longer need to maintain a set of visited nodes.
The predicate \texttt{GetForbiddenErrorsByPrecedence} maintains exactness of the Tesseract decoding algorithm, since there is still a unique path from $\emptyset$ to any set $F\in 2^{\mathscr{E}}$.
The more restrictive routine \texttt{GetForbiddenErrorsAtMostTwo} does not maintain exactness.
\begin{algorithm}[ht]
\caption{Pruning Predicate $P_{\mathrm{T}}(F, F')$}\label{alg:prune}
\SetKwFunction{Min}{Minimum}
\SetKwFunction{GetSoleElement}{GetSoleElement}
\SetKwFunction{GetForbiddenErrors}{GetForbiddenErrors}
\KwIn{$F, F' \in 2^{\mathscr{E}}$ where $(F'\supset F)\land(|F'| = |F|+1)$, and a function \GetForbiddenErrors: $2^\mathscr{E}\rightarrow 2^\mathscr{E}$}
\KwOut{$P_{\mathrm{T}}(F, F') \in \{0,1\}$}
$e \gets$ \GetSoleElement{$F'\setminus F$}\;
$J\gets $\GetForbiddenErrors{$F$}\;
\If{$e\in J$}{
    \Return $0$\;
}
$x \gets \mathscr{S} \oplus D(F)$\;
$d_{\min} \gets$ \Min{$x$}\;
\If{$e$ is incident to $d_{\min}$}{
    \Return $1$\;
}
\Return $0$\;
\end{algorithm}

\noindent{\bf A*:}
The A* pathfinding algorithm \cite{hart1968formal} changes the ordering of nodes in the priority queue by adding a {\it ``heuristic cost"} $h(F)$ to the cost of a node $F$.
Despite its name, $h(F)$ can be chosen in such a way that A* is still an exact algorithm. The property required is that the heuristic $h(F)$ give a strict lower bound on $d_G(F, \mathrm{EXIT})$. Such heuristics are called {\it admissible heuristics}.
In Algorithm~\ref{alg:Astarheuristic} we explain the heuristic used in Tesseract. It is simply a sum of evaluations of the \texttt{DetCost} routine, which is defined in Algorithm~\ref{alg:detcost}. This choice of heuristic is admissible, so Tesseract maintains its correctness guarantee.
One advantage of this routine is that for LDPC codes there is a constant amount of work needed to update the sum when a single error is added.
\begin{algorithm}[ht]
\caption{A* Heuristic Function $h(F)$}\label{alg:Astarheuristic}
\SetKwFunction{Min}{Minimum}
\SetKwFunction{GetForbiddenErrors}{GetForbiddenErrors}
\SetKwFunction{GetDetCost}{GetDetCost}
\KwIn{$F \in 2^{\mathscr{E}}$ and a function \GetForbiddenErrors: $2^\mathscr{E}\rightarrow 2^\mathscr{E}$}
\KwOut{$h(F) \in \mathbb{R}_{\geq 0}$}

$J\gets $\GetForbiddenErrors{$F$}\;
$x \gets \mathscr{S} \oplus D(F)$\;
$c\gets 0$\;

\For{$d \in x$}{
    $c \gets c \,+\, $\GetDetCost{$x, J, d$}\;
}

\Return $c$\;
\end{algorithm}

\begin{algorithm}[ht]
\caption{\texttt{GetDetCost}($x, J, d$)}\label{alg:detcost}
\SetKwFunction{Min}{Minimum}
\SetKwFunction{GetForbiddenErrors}{GetForbiddenErrors}
\SetKwFunction{GetDetCost}{GetDetCost}
\KwIn{A set of residual detection events $x\subset \mathscr{D}$, a set of forbidden errors $J\subset \mathscr{E}$, and an activated detector $d\in x$.}
\KwOut{\GetDetCost{$x, J, d$}$ \in \mathbb{R}_{\geq 0}$}
$c\gets \infty$\;
\For{$e \in E(d)$}{
    \If{$e\notin J$}{
      $c\gets $\Min$\left(c, \,\, \frac{w(e)}{|x\cap D(e)|}\right)$\;
    }
}
\Return $c$\;
\end{algorithm}

\noindent{\bf Beam search:}
The search through $G$ uses the A* pathfinding algorithm. To accelerate progress towards a solution we impose a {\it beam cutoff}.
For each node $F$ we visit, we compute the number of residual detection events $r(F) := |\mathscr{S}\oplus D(F)|$ and track the minimum value $r_{\mathrm{min}}$ of any visited node. Tesseract accepts a beam parameter \texttt{beam} and will not visit any node $F$ such that $r(F) > r_{\mathrm{min}}+ $\texttt{beam}.
A moderate beam of approximately 20 works well in practice. To avoid unbounded runtime and memory consumption, we also specify a maximum number \texttt{pqlimit} of nodes that can be added to the priority queue. After this many nodes are added, the Tesseract decoder terminates and declares a ``low-confidence" outcome. This is a heralded failure, but we treat all low-confidence outcomes as logical errors for the results in \sect{sec:results}.

\noindent{\bf Ensemble Reordering:} 
Algorithm~\ref{alg:prune} requires an absolute ordering of all detectors in $\mathscr{D}$.
We can improve the rate of convergence of the search by trying several different detector orderings. For the protocols we benchmarked, which have coordinate vectors in $\mathbb{R}^t$ assigned to each detector, the method we used to generate orderings is to sample a random normally distributed vector $\mathbf{z}\sim \mathcal{N}(0, 1)^t$  and order the detectors by the inner product of their coordinate vector with $\mathbf{z}$.
If multiple valid solutions are obtained, we output the minimum cost decoding as in \cite{shutty2024efficient}.

\noindent{\bf Beam Climbing:} Although a beam of $\approx 20$ is usually a good value, sometimes a larger or smaller beam works better. We found empirically that simply choosing a maximum beam value $B\in \mathbb{Z}_{\geq 0}$ and trying once for each beam value in the range $\{0, 1, \ldots, B\}$ works well.
This can be combined with ensemble reordering by using a different randomized total ordering on the detectors for each beam setting.

\noindent{\bf No-revisit detections:} With this heuristic, we maintain a set containing the data of $D(F)\oplus \mathscr{S}$ for each visited node $F$. After visiting the node $F$ we will then ignore (i.e., we do not visit) any nodes with the same leftover detection set.

\noindent{\bf Detection penalty:} This is a real number $c$ such that the ordering of the nodes in the priority queue is determined by the standard ordering plus a penalty term of $c\cdot |D(F) \oplus \mathscr{S}|$.
In other words, we add a cost of $c$ for each residual detection event. Similar to the beam cutoff, this discourages Tesseract from visiting nodes with many residual detection events (i.e., a high value of $r(F)$).

\section{Results}\label{sec:results}
\begin{figure}[ht]
    \centering
    \includegraphics[width=1.0\linewidth]{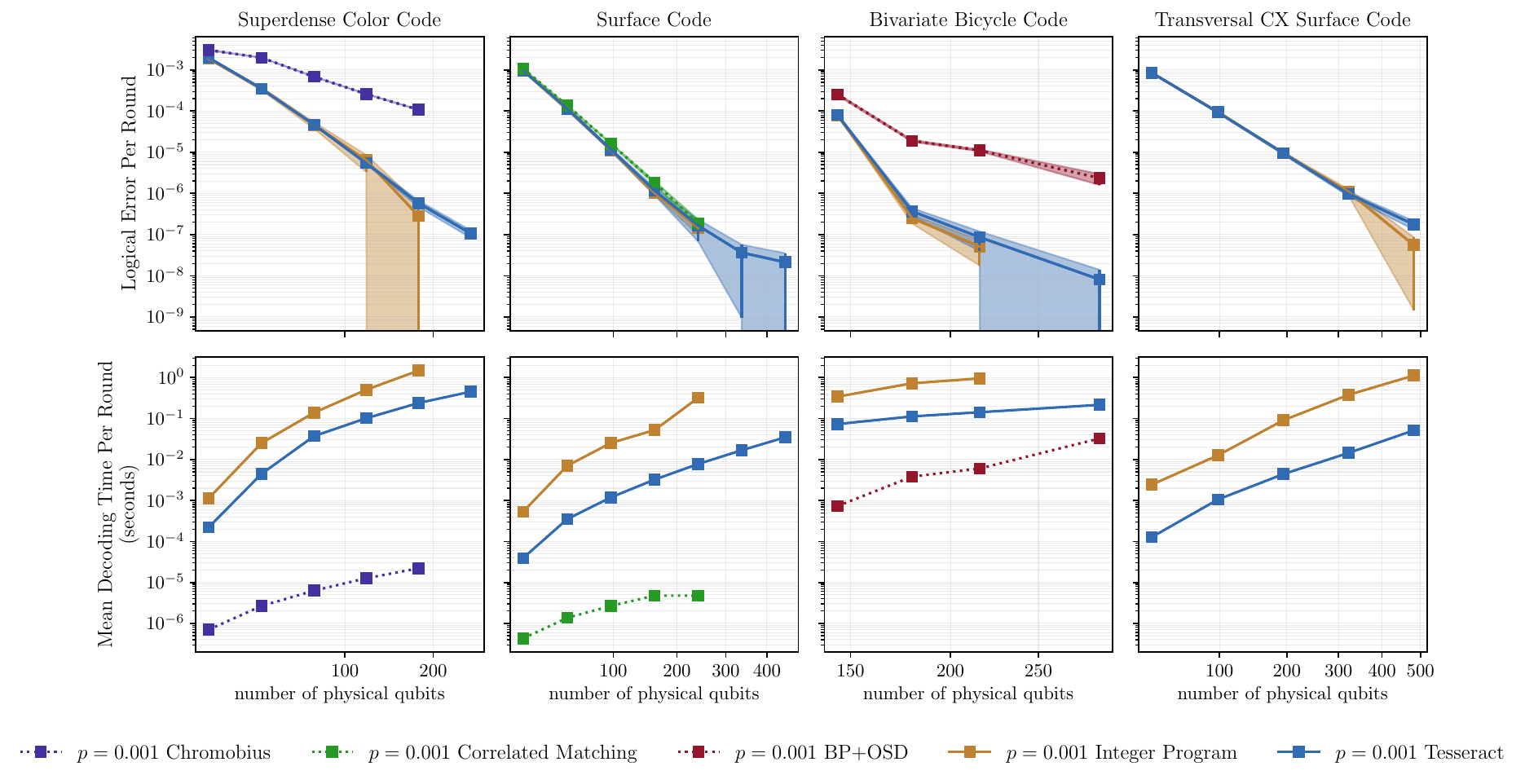}
    \caption{The Tesseract decoder achieves very similar accuracy to the Integer Program decoder while approximately 5$\times$ faster in runtime. Moreover, both Tesseract and the Integer Program decoders are order(s) of magnitude more accurate than fast algorithmic decoders on color codes (comparing with \texttt{chromobius} \cite{gidney2023new}) and bicycle codes (comparing with BP+OSD \cite{panteleev2021degenerate} via the \texttt{ldpc} module \cite{ldpc}). A square-root scaling is used for the number of physical qubits axis. Note that not all combinations of decoder, protocol, and physical error rate have an observed logical error, and these points are omitted from the upper row of plots. For example, the $d=13$ superdense color code circuit at $p=0.0001$ decoded with the Integer Program decoder had no observed logical errors within the scope of our simulations.} 
    \label{fig:protocolsTesseractSimplex}
\end{figure}
{\noindent {\bf Comparing Tesseract with other decoders:}} We benchmarked Tesseract against an integer programming decoder across four protocols: rotated surface code memories, superdense color code memories, transversal CNOT operations between rotated surface codes, and bivariate bicycle code memories (see Figure~\ref{fig:protocolsTesseractSimplex}).
At error rates of $p \leq 0.001$, Tesseract achieves decoding accuracy nearly identical to that of the integer program decoder for both of the topological codes, while operating approximately five times faster. At a higher error rate ($p = 0.002$), an error floor is observed for larger code distances (e.g., $d=11$ in the superdense color code), indicating a slight trade-off between speed and accuracy in this regime.
To the best of our knowledge, the performance of BP+OSD relative to near-optimal decoding for non-topological quantum LDPC codes has not previously been studied in detail.
We find that the logical error rate of Tesseract and the integer programming decoder are one to two orders of magnitude lower relative to the BP+OSD decoder. 
We use an uncorrelated (rather than a correlated) BP+OSD decoder, since uncorrelated BP+OSD has better accuracy (see Appendix \ref{sec:uncorrelated_bposd} for details).

To compute the logical error rate per round, we fix the error rate per shot $R_{\mathrm{per\, shot}} = \frac{\mathrm{number\, of\, errors}}{\mathrm{number\, of\, shots}}$ and the number of rounds $r$, and then use the equation
\begin{equation}
    R_{\mathrm{per\, round}} = \frac{1}{2}  \left(1 - (1 - 2 R)^{1 / r}\right).\label{eq:toPerRound}
\end{equation}
To find the error bars, we propagate a 90\% confidence interval through eq~\eqref{eq:toPerRound}.
For the transversal CX surface code circuits, we set $r$ to be the total number of rounds of syndrome extraction {\t across the two interacting surface codes}. This allows the error rate per round to be more directly compared with the surface code memory protocol.
For all other protocols, since they are just memory, we simply set $r$ to be the total number of rounds of syndrome extraction.

\begin{figure}[htbp]
    \centering
    \begin{subfigure}[t]{0.45\textwidth}
        \includegraphics[width=\textwidth]{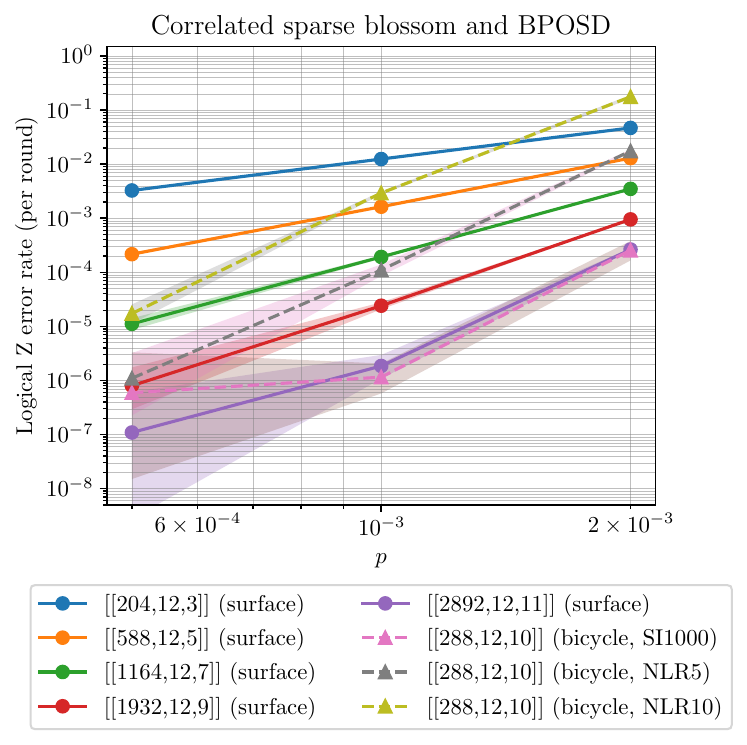}
        \caption{}
        \label{fig:q180_mwpm_bposd}
    \end{subfigure}
    \quad
    \begin{subfigure}[t]{0.45\textwidth}
        \includegraphics[width=\textwidth]{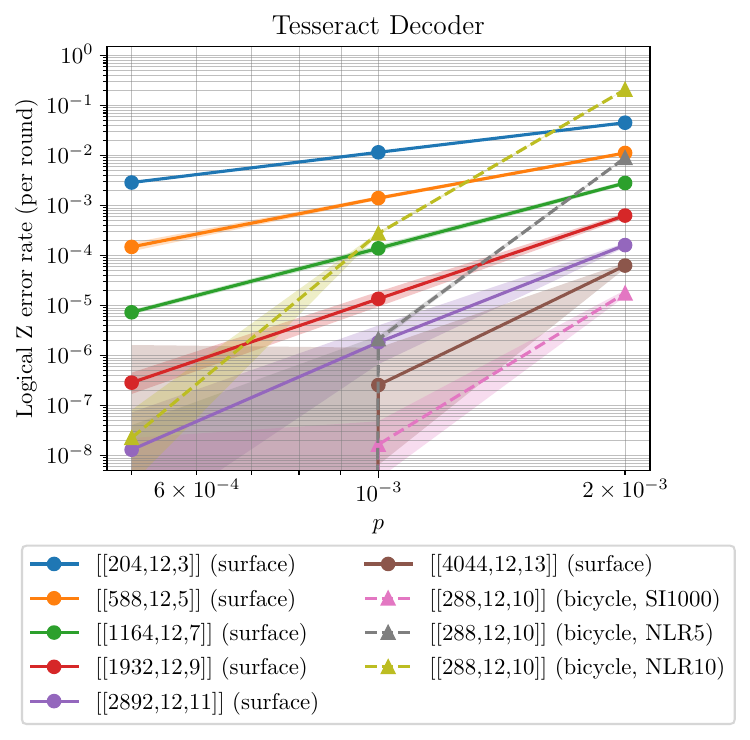}
        \caption{}
        \label{fig:q180_tesseract}
    \end{subfigure}
    \caption{We compare the performance of the [[144,12,12]] bivariate bicycle (BB) code from Ref.~\cite{bravyi2024high} with surface codes. The circuit for the [[144,12,12]] code uses 288 qubits including ancillas, and has circuit distance 10 (hence is referred to as $[[288,12,10]]$ in the figure). We compare with the performance of 12 copies of surface codes. In (a) we use correlated sparse blossom to decode the surface codes and BPOSD to decode the BB code, whereas in (b) we use our Tesseract decoder to decode both. We use an SI1000 noise model~\cite{Gidney2022benchmarkingplanar} for all surface code circuits and the BB code noise models are given in the legend. The ``NLR5'' and ``NLR10'' noise models use $5\times$ and $10\times$ higher noise strengths for couplers that are long-range on the toric layout.}
    \label{fig:surfaceVsBicycle}
\end{figure}

{\noindent {\bf Comparing the Bivariate Bicycle Code with the Surface Code:}} In Figure~\ref{fig:surfaceVsBicycle}, we compare surface codes to the [[144,12,12]] bivariate bicycle code (the ``gross code'') from Ref.~\cite{bravyi2024high}.
 The circuit for this code has 288 qubits including ancillas, and its circuit distance is at most 10 (hence its circuit parameters are [[288,12,10]]).
In Figure~\ref{fig:q180_mwpm_bposd} we use two-pass correlated sparse blossom \cite{higgott2025sparse} to decode the surface codes and BP+OSD \cite{panteleev2021degenerate,ldpc} to decode the bivariate bicycle code, whereas in Figure~\ref{fig:q180_tesseract} we use Tesseract to decode all circuits.
Using the less accurate decoders in Figure~\ref{fig:q180_mwpm_bposd}, we see that the $[[288,12,10]]$ bivariate bicycle (BB) circuit has similar performance to the distance 11 surface codes for SI1000 noise (a $10\times$ qubit saving, consistent with \cite{bravyi2024high}).
Intriguingly, we see that the relative performance of the $[[288,12,10]]$ BB circuit is much better using our Tesseract decoder, with the $[[288,12,10]]$ circuit having significantly better performance than distance 13 surface codes (a $14\times$ qubit saving), perhaps matching the performance of distance 15 surface codes (a $19\times$ qubit saving).
This demonstrates the advantage Tesseract offers in enabling a fairer assessment and comparison of QEC protocols, leading to a more accurate understanding of their relative performance.

We also use Tesseract to study how the performance of bicycle codes is impacted if long-range couplers are much noisier, which is perhaps a more reasonable noise model for 2D solid state architectures such as superconducting qubits.
Our results are shown in Figure~\ref{fig:surfaceVsBicycle}, where we compare a bivariate bicycle code with surface codes for a noise model where long-range couplers on a planar implementation of the BB code's toric layout (defined in~\cite{bravyi2024high}) have a higher noise strength.
A coupler is considered long-range if it is not an immediate neighbor on the toric layout, or if it would have to wrap around the torus (i.e.~long-range once boundaries are introduced).
For example, in the bulk of the bivariate bicycle code there are four short-range couplers and two long-range couplers.
We call these noisy long-range noise models ``NLR5'' and ``NLR10'', which use $5\times$ and $10\times$ higher noise strength for the two-qubit depolarizing channels after the long-range CZ gates, but are otherwise equivalent to an SI1000 noise model.
Using NLR10, at $p=0.1\%$ we find that the gross code (with circuit parameters [[288,12,10]]) is equivalent to distance 5 surface codes (rather than distance 11 for SI1000) using BP+OSD, and is equivalent to distance 7 surface codes (rather than distance 13 or 15) when using Tesseract.
In other words, qubit savings reduce from $10\times$ (SI1000) to $2\times$ (NLR10) using correlated matching and BP+OSD, and from $14\times$-$19\times$ (SI1000) to $4\times$ (NLR10) when using Tesseract.


\section{Comparison with \cite{ott2025decision}}
\begin{figure}[hbt!]
    \centering
    \includegraphics[width=0.7\linewidth]{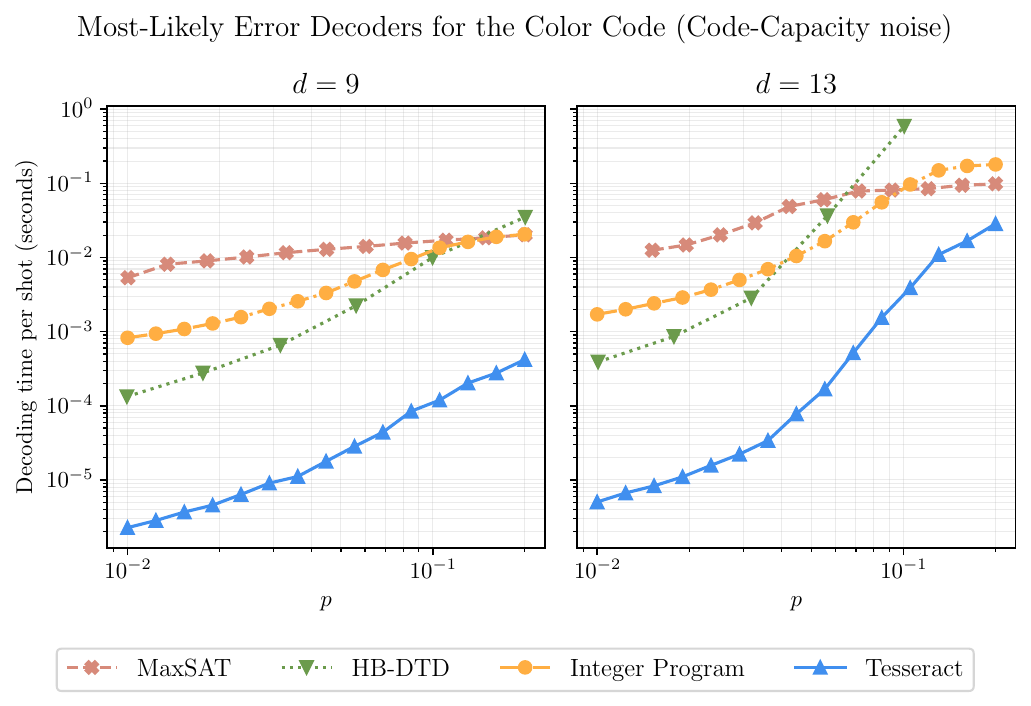}
    \caption{Comparison of the DTD and MaxSAT decoder timing data from \cite{ott2025decision} with our Integer Program and Tesseract decoder implementation. All of the above decoders are {\it exact} -- in particular, none of Tesseract's beam cutoffs were used -- guaranteeing that the most likely error is returned every time. It is worth noting that in practice, judicious use of the cutoffs such as Tesseract's beam parameter can make both Tesseract and the DTD decoder significantly faster without comprimising much accuracy.}
    \label{fig:dtdComparison}
\end{figure}

Tesseract is specialized for decoding quantum LDPC codes.
These include topological codes such as the surface code and color code, and other interesting codes such as bivariate bicycle (BB) codes \cite{dennis2002topological, bravyi2024high,bombin2006topological,baireuther2019neural,gidney2023new}.
There is tremendous interest in decoding algorithms for these codes.
Concurrent independent work shared in \cite{ott2025decision} introduced a similar idea to Tesseract.
They call it a ``Decision-Tree Decoder" (DTD). The DTD and Tesseract algorithms are closely related, as we discuss in the appendix.
Happily, there are complementary aspects of our work.
We explored different heuristic cutoffs with a greater emphasis on benchmarking with circuit-level noise models.
Ideally the insights from both of our works would be combined together to achieve the best performance.\footnote{For example, we expect the DTD decoder could be optimized by incorporating our canonicalized ordering of error paths, which removes redundancy from the search graph and avoids the need to track visited sets of errors.}
Tesseract is free open-source software written in high-performance C++, and our implementation appears to be somewhat faster than DTD.
We extracted the timing data from Figure 14 of \cite{ott2025decision} and made a comparison (Figure~\ref{fig:dtdComparison}).
Note we are unable to benchmark DTD directly on the same system so this is only a rough point of comparison.

\noindent {\bf Acknowledgments:}
We thank Michael Newman for suggesting simulating noisy long-range couplers. We thank Navin Kashyap, Benjamin Villalonga, Adam Zalcman, Cody Jones, Craig Gidney, Michael Newman, and Dripto Debroy for helpful conversations and feedback.

\bibliographystyle{alpha}
\bibliography{refs}

\appendix

\section{The challenge of handling $Y$ errors using BPOSD}\label{sec:uncorrelated_bposd}
\begin{figure}[htb!]
    \centering
    \includegraphics[width=0.5\linewidth]{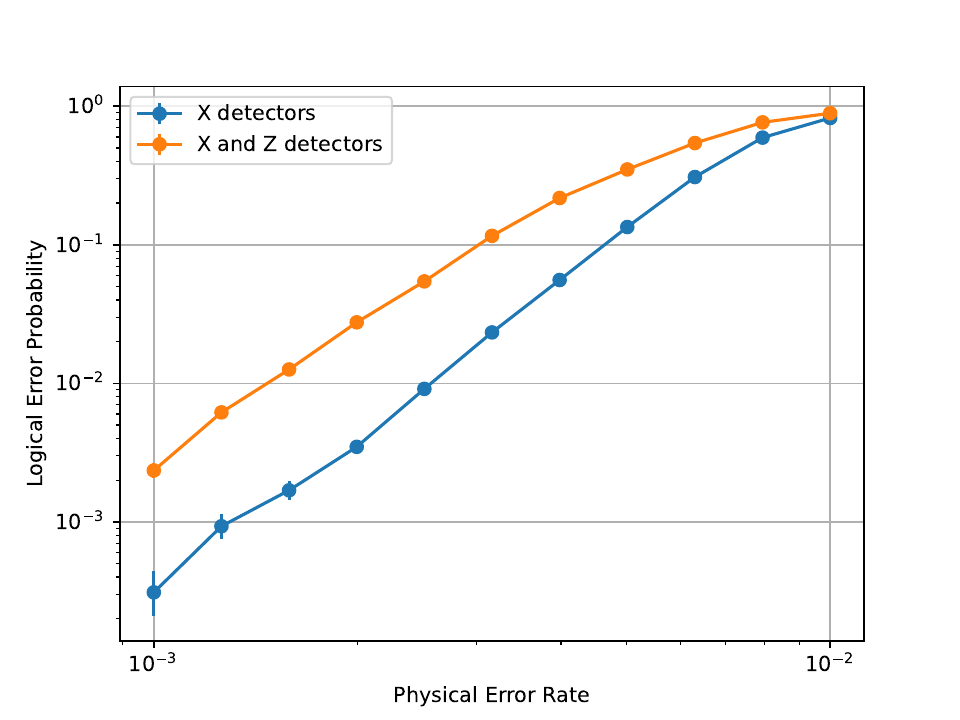}
    \caption{Comparison of uncorrelated vs.~correlated decoding using BP+OSD for the [[72,12,6]] bivariate bicycle code, using the same circuit and uniform circuit-level depolarizing noise model as given in Ref.~\cite{bravyi2024high}. We perform a 6-round $X$ memory experiment. For uncorrelated BP (labeled ``X detectors''), we decode a stim circuit with only the $X$-type detectors annotated, whereas for correlated BP+OSD we annotate all detectors ($X$-type and $Z$-type).}
    \label{fig:bposd_uncorrelated_vs_correlated}
\end{figure}

In this work, we use ``uncorrelated'' BPOSD, where we decode the $X$-type errors and $Z$-type errors separately.
We do this by annotating only the detectors of the same basis as the observable we are benchmarking (e.g.~only $X$ detectors, for an $X$ memory experiment).
Interestingly, we find that uncorrelated BPOSD is significantly more accurate than correlated BPOSD (in addition to being much faster, since it operates on a much smaller Tanner graph). See Figure~\ref{fig:bposd_uncorrelated_vs_correlated} for a comparison of the accuracy of both variants of BPOSD decoding for a [[72,8,6]] bivariate bicycle code circuit.
While this might initially seem counterintuitive, since the uncorrelated variant of BPOSD receives much less information about the error model, it can be understood by the fact that $Y$-type errors can cause trapping sets in BP-based decoders when both bases of detectors are annotated.
For example, if an $X$ stabilizer $S_X$ and a $Z$ stabilizer $S_Z$ overlap, they must do so on at least two qubits $i$ and $j$ to commute.
There is therefore necessarily a 4-cycle $(S_X,Y_i,S_Z,Y_j)$ in the full Tanner graph (detector error model) of a circuit implementing this code, if detectors in both bases are considered.
Furthermore, there are more sets of low-weight degenerate error configurations (e.g.~an $X$ and a $Z$ error on a qubit will have the same syndrome and a comparable probability to a $Y$ error).
Both degeneracy and short cycles in the Tanner graph are known to be problematic for BP-based decoders~\cite{poulin2008iterativedecoding}.
This issue motivates the development of decoders (such as Tesseract) that much better exploit $Y$ errors in circuits for quantum LDPC codes, and is one reason why our Tesseract decoder improves so significantly on BPOSD for the circuits we study here.

\section{Full results and benchmarking details}
\begin{figure}[!htb]
    \centering
    \includegraphics[width=1.0\linewidth]{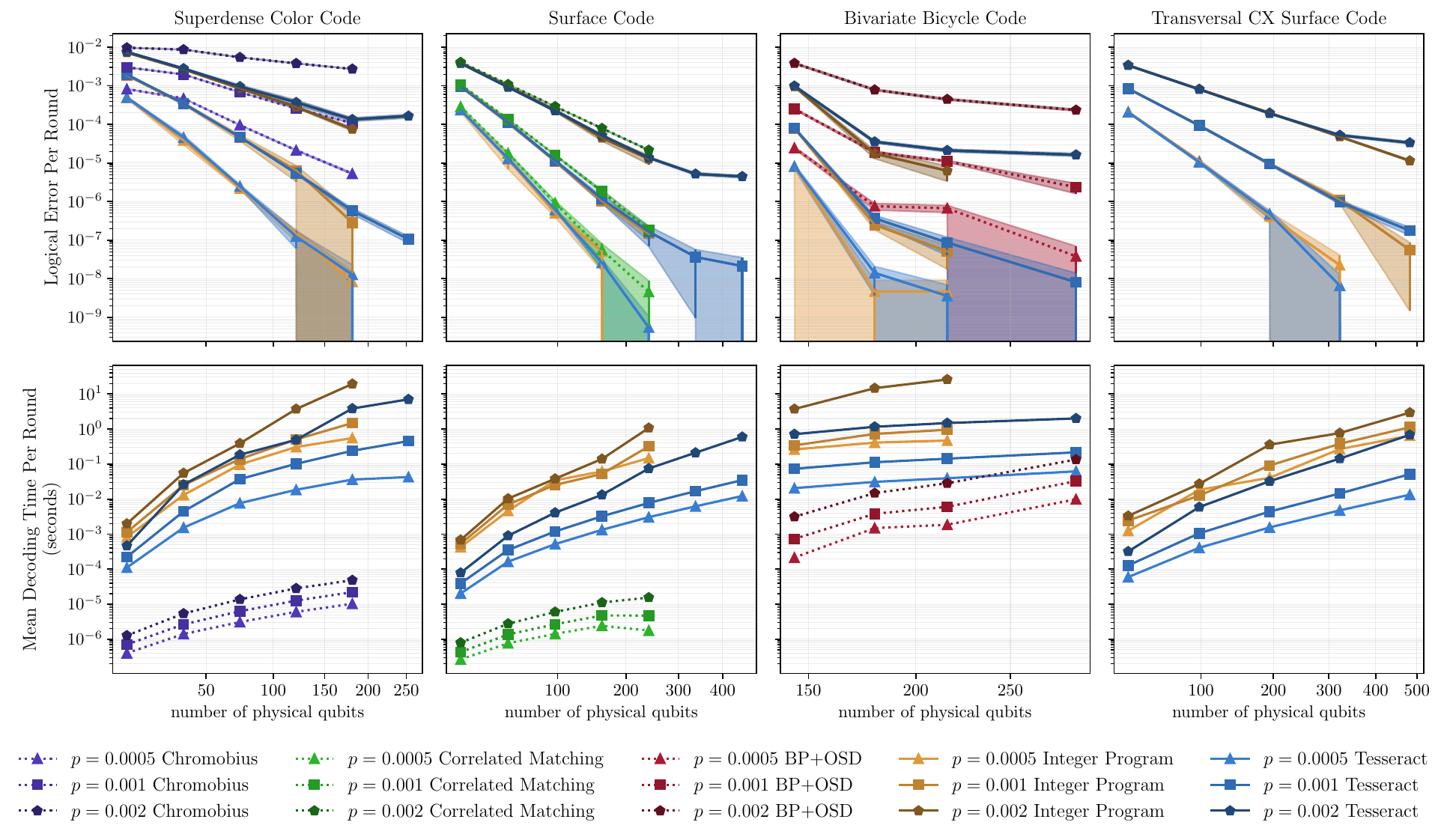}
    \caption{Results using Tesseract on a larger gamut of physical error rates: $p\in \{0.0005, 0.001, 0.002\}$.} 
    \label{fig:protocolsTesseractSimplexFull}
\end{figure}

We benchmarked Tesseract on SI1000 error rates $p\in \{0.0005, 0.001, 0.002\}$. The full results are shown in Figure~\ref{fig:protocolsTesseractSimplexFull}.
We used two parameter settings in our benchmarking.
The {\it short beam} setting is a beam of 15 with beam climbing combined with an ensemble of 16 different detector orderings, and a \texttt{pqlimit} of 200,000. The {\it long beam} setting is a beam of 20 with beam climbing combined with an ensemble of 21 different detector orderings, and a \texttt{pqlimit} of 1,000,000. We used the long beam for these protocols:
\begin{enumerate}
    \item All surface code transversal CX protocols.
    \item The superdense color code at these distance and error rate combinations: $(d,p)\in \{11, 13\}\times \{0.001, 0.002\}$.
    \item The surface code at these distance and error rate combinations: $(d,p)\in \{11,13\}\times \{0.002\}$.
\end{enumerate}
We used the short beam for all other protocols, including all bicycle codes. We enabled the no-revisit detections heuristic for all protocols. We did not use a detection penalty.
In future work, it would be helpful to automate the selection of beam parameters.

\noindent{\bf System Information: } We compiled Tesseract using Clang 18.1.8 (Red Hat 18.1.8-1) on a 64-bit x86\_64 Linux system. The benchmarks were conducted on a machine equipped with two Intel\textsuperscript{\textregistered} Xeon\textsuperscript{\textregistered} CPUs running at 3.10 GHz, with a total of 60 logical processors (15 physical cores per CPU, 2 threads per core). Each decoder was executed on a single physical core, but no explicit resource isolation was enforced to control system load. As a result, the reported execution times are intended to be representative estimates rather than precise measurements under controlled conditions.

\section{Technical differences from \cite{ott2025decision}}
We will explain a few of the similarities and differences between Tesseract and DTD at a high level.
First, it is important to note that both our work and \cite{ott2025decision} each provide {\it two decoders}:
\begin{enumerate}
    \item A ``slower decoder" that has a rigorous guarantee of optimality
    \item A ``faster decoder" that sacrifices some amount of accuracy for improved performance
\end{enumerate}
In our open-source implementation, the command-line arguments can be adjusted to interpolate between these modes.
In both algorithms, the A* search procedure is used.\footnote{Although the authors of \cite{ott2025decision} do not identify it as such, what they term the ``syndrome height" is the same as the  ``admissible heuristic" used in A*.}
Tesseract traverses the graph of error sets slightly differently with a canonicalized path ordering.
There are also differences in the heuristics or ``cutoffs" used to improve performance. In \cite{ott2025decision}, the authors 
explore more sophisticated admissible heuristics for special cases such as $k$-colorable graphs. They also consider BP-guided search. Tesseract uses a simpler A* heuristic cost calculation and a beam cutoff. We both make use of ensembling techniques.
We also benchmark our decoders differently.
In \cite{ott2025decision} there is some emphasis on code-capacity noise models which assume noise-free measurement. These error models are of fundamental interest, but do not give a representative picture of practical runtimes on circuit-level noise.
Our focus is on applying Tesseract to circuit-level noise models for color codes, surface codes, bicycle codes, and transversal CNOT operations between two surface codes, applicable to neutral atom architectures \cite{cain2024correlated}.
We also focus on the high-accuracy regime. We achieve 100x lower logical error rates than BP+OSD for the bicycle codes at $p=0.001$, suggesting that the ``fast decoder" (BP+DTD) of \cite{ott2025decision} is leaving a factor of about 10x in logical error rate on the table (see Fig. 12 of \cite{ott2025decision}).
Of course, this is not to say that the DTD cannot achieve the same accuracy, but rather that with BP+DTD they probed a different operating regime where the algorithm is significantly faster and less accurate.
Lastly, as mentioned before, our implementation appears significantly faster for the case of exact decoding of color codes under code capacity noise.
Together, our results provide strong motivation for future work improving and applying search-based decoders for QEC codes.

\section{Additional results comparing surface codes and bivariate bicycle codes}

\begin{figure}
    \centering
    \includegraphics[width=1.0\linewidth]{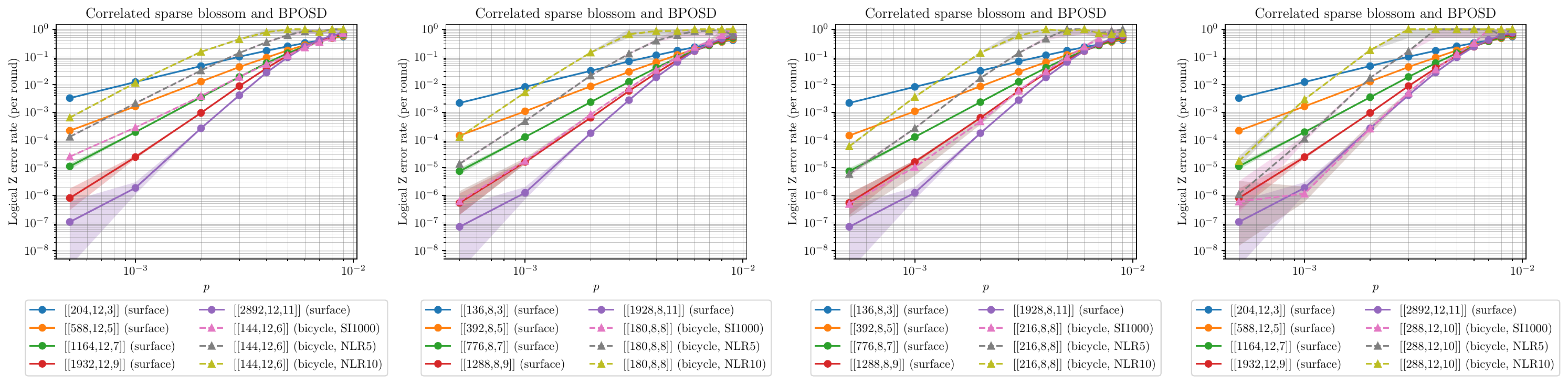}
    \caption{A comparison of surface codes and bivariate bicycle codes. Here the surface codes are decoded with correlated sparse blossom and the bivariate bicycle codes are decoded with BP+OSD. All surface code circuits use an SI1000 noise model and we use $k$ copies of surface codes to compare with a bicycle code encoding $k$ logical qubits. The bivariate bicycle codes use SI1000, NLR5 and NLR10 noise models (given in the legend).}
    \label{fig:all_bicycle_vs_surface_bposd_matching}
\end{figure}

\begin{figure}
    \centering
    \includegraphics[width=1.0\linewidth]{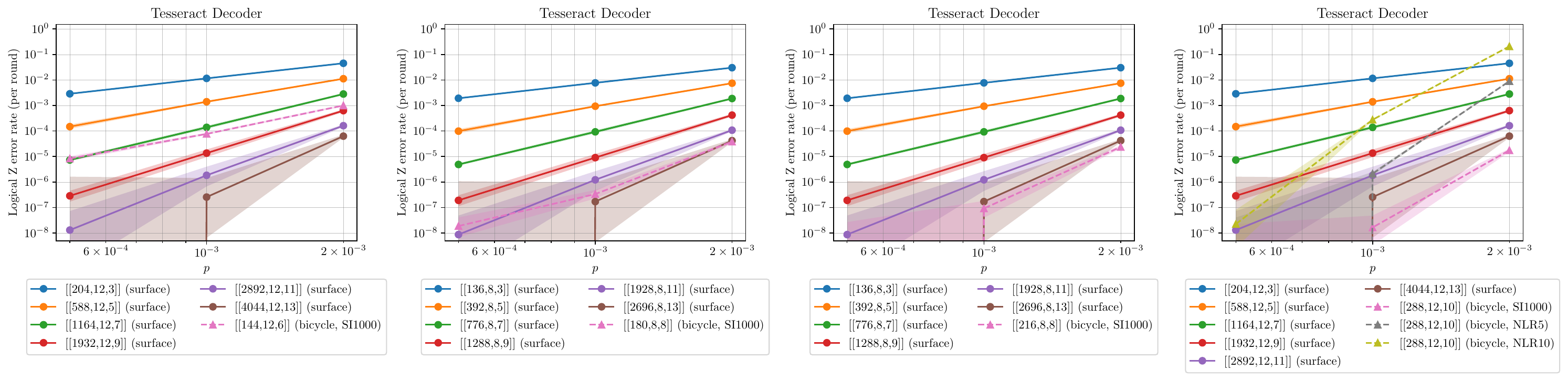}
    \caption{A comparison of surface codes and bivariate bicycle codes, all decoded with Tesseract. All surface code circuits use an SI1000 noise model and we use $k$ copies of surface codes to compare with a bicycle code encoding $k$ logical qubits. The bivariate bicycle codes use SI1000, NLR5 and NLR10 noise models (given in the legend).}
\label{fig:all_bicycle_vs_surface_tesseract}
\end{figure}

In this section we present additional results comparing surface codes to the four smallest bivariate bicycle codes from \cite{bravyi2024high}.
In Figure~\ref{fig:all_bicycle_vs_surface_bposd_matching} we use correlated matching to decode the surface codes and BP+OSD for the bivariate bicycle codes, whereas in Figure~\ref{fig:all_bicycle_vs_surface_tesseract} we use Tesseract to decode all codes.
We verified with a MaxSAT solver\footnote{We used the  \texttt{Circuit.shortest\_error\_sat\_problem}  method in \texttt{stim} \cite{stim} combined with a solver from the MaxSAT Evaluation 2024 \cite{berg2024maxsat}.} that the circuit distance of the [[72,12,6]] code is 6 (circuit parameters [[144,12,6]]) and the circuit distance of the [[90,8,10]] code is 8 (circuit parameters [[180,8,8]]).
For the larger codes we give the upper bound on the circuit distance given in \cite{bravyi2024high}.
Note that we use $d$ rounds of measurements for all bivariate bicycle code circuits, where $d$ is the distance of the code.

\end{document}